\documentclass[aps,groupedaddress]{revtex4}
\usepackage{epsfig}
\def\numdim{N}
\def\nummarked{M}
\def\maxdim{\numdim_{max}}
\def\bigo{O}
\def\ket#1{\left\vert #1 \right\rangle}
\def\bra#1{\left\langle #1 \right\vert}
\def\oracle{O}
\def\ox{\mathbf{x}}
\def\ofunc{f}
\def\identity{I}
\def\error{\epsilon}
\def\prob{p}
\def\diracd{\delta}
\def\ensavg#1{\left\langle#1\right\rangle}
\def\grover{G}
\def\unitary{U}
\def\oparrow#1{\buildrel \rm #1 \over \rightarrow}
\def\nv{n}
\def\nvb{\hat{\nv}}
\def\pauli{\sigma}
\def\paulib{\vec{\pauli}}
\def\rot{R}
\def\abs#1{\left\vert #1 \right\vert}

\def\bkm#1#2{\left\langle #1 \vert #2 \right\rangle}

\def\refsec#1{Sec.\ \ref{Section::#1}}

\def\refeqn#1{Eq.\ (\ref{Equation::#1})}

\def\refeqs#1#2{Eqs.\ (\ref{Equation::#1}) and (\ref{Equation::#2})}
\def\refeqto#1#2{Eqs.\ (\ref{Equation::#1}--\ref{Equation::#2})}

\begin{document}


\title{Effects of Random Noisy Oracle on Search Algorithm Complexity}

\author{Neil Shenvi}
\author{Kenneth R. Brown}   
\author{K. Birgitta Whaley}
\affiliation{Department of Chemistry and the Kenneth S. Pitzer Center for Theoretical Chemistry, University of California, Berkeley,
        Berkeley, CA 94720}

\date{\today}

\begin{abstract}

Grover's algorithm provides a quadratic speed-up over classical 
algorithms for unstructured database or library searches.  This paper 
examines the robustness of Grover's search algorithm 
to a random phase error in the oracle and analyzes the complexity of the 
search process as a function of the scaling of the oracle error with 
database or library size. Both the discrete- and continuous-time 
implementations of the search algorithm are investigated.  It is shown 
that unless the oracle phase error scales as $O(N^{-1/4})$, neither 
the discrete- nor the continuous-time implementation 
of Grover's algorithm is scalably robust to this error in the absence of 
error correction.
\end{abstract}


\maketitle

\setcounter{figure}{0}

\section{Introduction} \label{Section::Introduction}
Issues of fault tolerance and error correction are of both theoretical and 
practical interest in designing and implementing quantum algorithms.
One approach to diminishing the effects of error focuses on error correcting 
codes, which can be used to correct errors once they have occurred 
\cite{Shor:95,Preskill:98}.  
Another approach which has also proven successful is that of decoherence-free 
subspaces, which involves mapping a computation
onto a subspace that is relatively free from error \cite{Bacon:00}.  This 
second 
approach is of interest because it is an example of passive, rather than active, error 
correction.  The success of the passive approach leads to the question of whether 
existing quantum algorithms are inherently robust to errors, and, if they are 
not, whether there might exist modified implementations of these algorithms that
are robust to certain errors, i.e. without requiring active error correction.  

In this paper, we analyze the robustness of Grover's algorithm to error.  Grover's 
search algorithm is the basis for a number of quantum algorithms \cite{Abrams, Novak, Brassard}.
Most of these algorithms involve the use of an oracle, a black box 
device which takes as input a quantum state in the computational 
basis 
and returns as 
output 
some function of that state.  
Given a perfect, noise-less oracle, Grover's search algorithm attains a quadratic 
speedup over classical search algorithms.  However, this speed-up is predicated upon the 
perfect implementation of the oracle.  Although, for the purposes of analysis, the oracle 
is often treated as a ``black box'' whose inner workings are unknown, 
any physical implementation of Grover's algorithm must also include a 
physical implementation of the 
oracle and this may be imperfect.  Therefore it is of interest to 
ascertain what effect the accuracy of the oracle 
implementation has on the overall complexity of the algorithm.  We make use of this 
term here as it is employed in computer science terminology, namely, the complexity 
of an algorithm is defined as the number of computational steps required in order to 
achieve a pre-determined fixed probability of success.

Several previous papers have studied the effects of oracle noise on 
Grover's algorithm, 
using various models~\cite{Long:99, Hsieh:02, Pablo-Norman:99, Kwon:02, 
Biham:03}.  We 
consider here the 
random phase error model, addressing it within both discrete- and 
continuous-time implementations of the search algorithm. The effect of 
random phase errors on the discrete-time Grover algorithm was already 
studied numerically in \cite{Long:99}.  In this 
paper we derive analytic results for this model and present numerical 
evidence verifying the validity of these results. We analyze the 
complexity of the search algorithm as a function of the scaling of the 
errors, and arrive at bounds on the error that must be satisfied for a 
constant probability of success.  
In particular, we find that in order to achieve a constant success 
probability independent of the library size that is being searched, the 
oracle error must scale at most as $N^{-1/4}$ where $N$ is the library size. 
We also analyze the effect of phase 
errors on a continuous-time analogue of the search algorithm.  This is 
found to be relatively straightforward to study within a density matrix 
approach.  We find that this leads to similar results for the 
complexity 
as a function of the scaling of the errors, 
namely that there is an upper bound of $\bigo(N^{1/2})$ on algorithmic 
complexity for $\delta 
\geq 1/4$ and $\bigo(N^{1-2\delta})$ for $\delta \leq 1/4$, when the error 
scaling goes as $N^{-\delta}$.  We further show that these 
bounds are tight; in other words, that the algorithmic complexity has a 
lower bound of $\Omega(N^{1/2})$ for $\delta \geq 1/4$ and 
$\Omega(N^{1-2\delta})$ for $\delta \leq 1/4$.  (Following standard 
computer science 
notation, we will use the following to characterize the growth of certain 
functions:  We will say that $f(n)$ is bounded above by $g(n)$, {\it i.e.} 
that $f(n) = 
\bigo(g(n))$, if there are positive constants $c$ and $k$ such that $0 
\leq f(n) \leq 
c g(n)$ for $n \geq k$.  Similarly $f(n)$ is bounded below by $g(n)$, 
{\it i.e.} $f(n) = \Omega(g(n))$, if $0 \leq c g(n) \leq f(n)$ 
for constants $c,k \geq 0$ and $n \geq k$.)

This result has practical importance in determining the amount of 
oracle error that is allowable as the size of the library is increased.  
We will show that our complexity results imply that if the library size 
is increased by a factor of $k$, the oracle error must decrease 
by a factor of $k^{1/4}$ in order to attain a similar probability of 
success.

The remainder of this paper is organized as follows.  
\refsec{DiscreteTime} describes the random phase error model for a noisy 
oracle, summarizes the Grover search algorithm in a discrete-time 
implementation, and then derives the effect of the oracle noise on this 
implementation.  \refsec{ContinuousTime} derives the effect 
of oracle noise on the continuous-time formulation of Grover's algorithm 
proposed by Farhi et al. in \cite{Farhi:98,Farhi:00}.  We summarize and 
present conclusions in \refsec{Conclusions}.

\section{Discrete-Time Search} \label{Section::DiscreteTime}
\subsection{Quantum Search with a Phase Oracle} \label{subsection:oracle_search}

A phase oracle is a quantum oracle that ``marks'' 
one or more computational basis states with a specified phase 
(usually $-1$).  
For a function $\ofunc(\ox) \to \{0,1\}$, where $\ox$ denotes an n-bit binary string, a phase oracle implements the following
operation:
\begin{equation} \label{Equation::NoiselessOracle}
\ket{\ox} \oparrow{\oracle} (-1)^{\ofunc(\ox)}\ket{\ox}
\end{equation}
The search problem is phrased as follows:
Given an oracle, $\oracle$, which implements the function $\ofunc(\ox) \to 
\{0,1\}$,
find a state in the computational basis $\ox$ for which $\ofunc(\ox) 
= 1$.  Assuming a 
perfect, 
noiseless oracle 
(the concept of a noisy oracle is introduced below), the well-known 
result for the complexity of Grover's algorithm, {\it i.e.}, for the 
quantum search problem, is $\bigo(\sqrt{\frac{\numdim}{\nummarked}})$, 
where $\nummarked$ is the number of states for which 
$\ofunc(\ox) = 1$~\cite{Grover:96, Grover:97, Bennett:2, Boyer}.  In contrast, the classical complexity of the search problem is 
$\bigo(\frac{\numdim}{\nummarked})$~\cite{Knuth:book}.  Thus, 
the quantum algorithm provides a quadratic speed-up over the classical algorithm.

\subsection{Grover's algorithm in absence of noise} 
\label{subsection:nonoise}

The description of Grover's algorithm in this paper follows the discussion 
in \cite{Nielsen:00}.  The search is made on the set of $N = 2^n$ states 
represented by all n-bit binary strings $x \in \{0,1\}^n$.  The states are 
denoted by $\ket{x}$.  Within a discrete-time computation model, Grover's 
algorithm searches for marked or ``target'' states via repeated 
application of the Grover iteration operator, $\grover$, 
which can be written as:
\begin{equation}
\grover = \left(2\ket{\eta}\bra{\eta}-\identity\right)\oracle
\label{Equation::iterate}
\end{equation}
where $\ket{\eta} = \frac{1}{\sqrt{\numdim}}\sum_{\ox=0}^{\numdim-1}{\ket{\ox}}$ is the
equal superposition over all states.  From this point on, we will assume 
that $\nummarked = 1$ to simplify the discussion.  The arguments can 
easily be extended to the case where $\nummarked > 1$.  Let the 
state $\ket{\tau}$ be the ``marked'' 
state.  Then we can write out the explicit form of the oracle as:
\begin{equation} \label{Equation::Oracle}
\oracle = \identity + \left(e^{i\pi}-1\right)\ket{\tau}\bra{\tau}
\end{equation}

Assuming that our registers are initialized
to $\ket{\eta}$, it can be shown that after $\bigo(\sqrt{\numdim})$ applications of 
$\grover$, the quantum computer is approximately in the marked state 
$\ket{\tau}$ 
\cite{Grover:96, 
Grover:97}.  This result is demonstrated by noting that the search 
algorithm can be mapped onto a two-state
subspace spanned by the two basis vectors $\ket{1}$ and $\ket{2}$, 
where
\begin{eqnarray}
\label{Equation::BasisKetA}
\ket{1} &=& \ket{\tau}, \\
\label{Equation::BasisKetB}
\ket{2} &=& \frac{\ket{\eta}-\ket{\tau}\bkm{\tau}{\eta}}{\sqrt{1-\abs{\bkm{\tau}{\eta}}^2}}.\end{eqnarray}
In other words, $\ket{1}$ is the marked state and $\ket{2}$ is the 
equal superposition over all unmarked states.

When our initial state, $\ket{\eta}$, and the operator $\grover$ 
are rewritten in the $\ket{1}$, $\ket{2}$ basis,
\begin{equation}
\ket{\eta} = \left( \matrix{
 \sqrt{\frac{1}{\numdim}} \cr
\sqrt{\frac{\numdim-1}{\numdim}} \cr} 
\right) 
\end{equation} 
\begin{equation}
\grover = \left( \matrix{ 1-\frac{2}{\numdim} & \frac{2
\sqrt{\numdim-1}}{\numdim} \cr -\frac{2 \sqrt{\numdim-1}}{\numdim} &
1-\frac{2}{\numdim} \cr} \right) = \left( \matrix{ \cos(\Theta/2) &
\sin(\Theta/2) \cr -\sin(\Theta/2) & \cos(\Theta/2) \cr} \right),
\end{equation}
where 
\begin{equation}
\sin(\Theta/2) = \frac{2\sqrt{\numdim-1}}{\numdim},
\label{Equation:noiseless}
\end{equation}
we see that the effect of repeated applications of $\grover$ is to make 
successive rotations by $\Theta/2$ on state $\ket{\eta}$.  This 
convention 
for the definition of $\Theta$ is useful because $\Theta$ represents 
the angle of rotation applied by the operator $\grover$ on the Bloch 
sphere defined for the Grover subspace spanned by the basis states 
$\ket{1}$ and $\ket{2}$.
We can therefore view Grover's algorithm as the process of rotating our initial state,
$\ket{\eta}$, onto the target state, $\ket{\tau} \equiv \ket{1}$ by a discrete sequence of small rotations. For small $\Theta$ (i.e.
large $\numdim$), $\sin\Theta \approx \Theta$, yielding an incremental 
rotation angle of
approximately $\frac{4\sqrt{\numdim-1}}{\numdim} \approx
\frac{4}{\sqrt{\numdim}}$.  Then a rotation of $\pi$ radians on the Bloch 
sphere requires $\bigo(\sqrt{\numdim})$ applications of $\grover$.  Hence, 
Grover's search algorithm requires $\bigo(\sqrt{\numdim})$ calls to the 
oracle.  The well-known lower bound for quantum search has been 
established as $\Omega(\sqrt{\numdim})$ 
\cite{Bennett:97}.  Thus, Grover's algorithm is an optimal search.

\subsection{Grover's algorithm in presence of oracle noise} 
\label{subsection:noisy}
This $\bigo(\sqrt{\numdim})$ bound on the running time of the
search assumes that the oracle exactly implements the phase operation
specified by \refeqn{NoiselessOracle}.  Instead, we could envision a more 
realistic
oracle which implements \refeqn{NoiselessOracle} with some finite degree 
of precision. Specifically, we adopt here the concept of a noisy phase
oracle introduced in \cite{Long:99} which implements the following operation: 
\begin{equation}
\label{Equation::NoisyOracle} 
\ket{\ox} \oparrow{\oracle} \left(e^{i(\pi+\error)}\right)^{\ofunc(\ox)}\ket{\ox}, 
\end{equation} 
where $\error$
is a small, random phase error ($\error \ll \pi$) with probability
distribution $\prob(\error)$  \footnote{For other models of oracle noise, see 
\cite{Long:99, Pablo-Norman:99, Hsieh:02}.}.  We will make the assumption 
that the error is 
zero-mean, i.e. that $\ensavg{\error} = 0$.  For $\prob(\error) = 
\diracd(\error)$, the noiseless oracle is recovered.  In practice, the 
root mean-square magnitude of 
$\error$ can be made small through a careful physical implementation of the oracle.  
However, due to the finite precision of any experimental implementation, 
the average {\em magnitude} of $\error$ will never be zero, {\it i.e.}, $\epsilon_{rms}=\sqrt{\ensavg{{\error}^2}} > 0$.

Given this more realistic oracle model, we now investigate whether Grover's search 
algorithm is still $\bigo(\sqrt{\numdim})$.  The explicit form of the 
noisy oracle can be written as,

\begin{equation}
\oracle = \identity + (e^{i(\pi+\error)}-1)\ket{\tau}\bra{\tau}
\end{equation}
Then Grover's iteration operator, $\grover$, becomes:
\begin{equation} \label{Equation::NoisyGrover}
\grover = \left( \matrix{
(-1+\frac{2}{\numdim})e^{i(\pi+\error)} & \frac{2 
\sqrt{\numdim-1}}{\numdim} \cr
\frac{2 \sqrt{\numdim-1}}{\numdim}e^{i(\pi+\error)} & 1-\frac{2}{\numdim} \cr}
\right),
\end{equation}
which is clearly still unitary.
To see the effects of the random error, we first note that any single 
qubit unitary operator can be written as
\begin{equation} \label{Equation::UnitaryExpansion}
\unitary = 
\exp(i\alpha)\rot_{\nvb}(\Phi) 
= \exp(i\alpha)\left(\cos{\left(\frac{\Phi}{2}\right)}\identity 
- i\sin{\left(\frac{\Phi}{2}\right)}\nvb\cdot\paulib\right),
\end{equation}
where $\paulib$ are the Pauli operators.
Thus, the Grover's iteration operator using the 
noisy oracle is equivalent, up to an overall global phase factor, to a 
rotation, $R_{\nvb} (\Phi)$, on the Bloch sphere by some angle $\Phi$ 
about some direction $\nvb$.  
We can determine the value of $\Phi$ and $\nvb$ by using \refeqn{NoisyGrover} and 
\refeqn{UnitaryExpansion},
\begin{eqnarray} 
\label{Equation::GroverParamsA}
\cos{\left(\frac{\Phi}{2}\right)} & = & 
\left(1-\frac{2}{\numdim}\right)\cos{\left(\frac{\error}{2}\right)}\\
\label{Equation::GroverParamsB}
\sin{\left(\frac{\Phi}{2}\right)}\nv_x & = &
\frac{2\sqrt{\numdim-1}}{\numdim}\sin{\left(\frac{\error}{2}\right)}\\
\label{Equation::GroverParamsC}
\sin{\left(\frac{\Phi}{2}\right)}\nv_y & = &
-\frac{2\sqrt{\numdim-1}}{\numdim}\cos{\left(\frac{\error}{2}\right)}\\
\label{Equation::GroverParamsD}
\sin{\left(\frac{\Phi}{2}\right)}\nv_z & = &
-\left(1-\frac{2}{\numdim}\right)\sin{\left(\frac{\error}{2}\right)}.
\end{eqnarray}
A noiseless oracle can be recovered using these equations by 
setting $\error$ to zero.  In that case we obtain the large $N$ Grover rotation angle, $\Phi \approx \frac{4}{\sqrt{\numdim}}$, with $\nv_x = \nv_z = 0$, and $\nv_y = 1$.  Hence, the 
noiseless Grover's algorithm corresponds to a rotation on the Bloch sphere about the $y$-axis. The effect of a noisy oracle is to add small, random perturbations 
to this rotation axis, thereby changing the pure $y$-axis rotation 
to a rotation about an axis $\nvb$ which also contains non-zero $x$- and $z$- components.  

To analyze the running time of the noisy search algorithm, we will take the 
limit of large $\numdim$.  This is a useful assumption for our current purposes since we are interested in 
the complexity of the search algorithm for large $\numdim$.
From \refeqto{GroverParamsA}{GroverParamsD}, we can express $\grover$ up to a 
global phase 
factor as,
\begin{equation}
\grover = 
\exp\left(-i\Phi\left(\left(-1+\frac{2}{\numdim}\right)\sin{\left(\frac{\error}{2}\right)}\pauli_z 
+ 
\frac{2\sqrt{\numdim-1}}{\numdim}\left(\sin{\left(\frac{\error}{2}\right)}\pauli_x 
- 
\cos{\left(\frac{\error}{2}\right)}\pauli_y\right)\right)/\sin{\left(\frac{\Phi}{2}\right)}\right),
\end{equation}
where $\sin(\frac{\Phi}{2}) = 
\sqrt{1-(1-\frac{2}{\numdim})^2\cos^2(\frac{\error}{2})}$.
Since both $\error$ and $\numdim^{-1/2}$ are small parameters, we can use a 
double Taylor expansion of the terms in the exponent with respect to 
$\error$ and $\numdim^{-1/2}$.  Excluding second-order terms 
and higher, we 
obtain
\begin{equation}
\grover = 
\exp\left(i \left(\error \pauli_z + 
\frac{4}{\sqrt{N}}\pauli_y\right)\right).
\end{equation}

Again using the fact that $\error$ and $\numdim^{-1/2}$ are 
both small parameters, 
we can make use of the Baker-Campbell-Hausdorff formula~\cite{messiahref} to arrive at
\begin{equation} \label{Equation::Grover2R}
\begin{array}{rcl}
\grover &=& 
\exp\left(i\error\pauli_z\right)\exp\left(i\frac{4}{\sqrt{\numdim}}\pauli_y\right) 
+ \bigo(\frac{\error}{\sqrt{\numdim}}) \\
&\approx&
\rot_{\hat{z}}(-\error)\rot_{\hat{y}}(-\frac{4}{\sqrt{\numdim}}).
\end{array}
\end{equation}

Examining \refeqn{Grover2R}, we observe that there are two important 
timescales that will be relevant to any physical implementation.  
The first is the timescale of the $y$-axis 
rotation, 
$\sqrt{\numdim}$. 
 Taking the limit as $\error \to 0$, we see 
that the $y$-axis rotation, $R_{\hat{y}}$ is the ``driving force'' of 
the search algorithm, 
in that it rotates the initial state, $\ket{\eta}$, towards the target state, 
$\ket{\tau}$.
The second important timescale, $1/\error$, is the timescale of the 
random rotation about the z-axis of the Bloch sphere, 
$R_{\hat{z}}$. This rotation is the effect of noise and does not advance the search process.  Because $\error$ is a random 
variable, this rotation about $z$ will in general be different for each application of $\grover$.  
However, $\error$ does have a characteristic root-mean-square magnitude,
$\error_{rms}$, which is 
determined by the probability 
distribution $\prob(\error)$.  Then the characteristic timescale of
the $z$-axis rotation is determined by $1/\error_{rms}$.

In general, $\error_{rms}$ and $\sqrt{\numdim}$ are independent 
parameters: $\error_{rms}$ is the phase error that is specific to the experimental 
implementation, while $N$ is the size of the database.  However, in 
reality, these two parameters will be linked for a given experimental 
setup because the size of the database ({\it i.e.}, the number of states 
$N=2^n$ accessed by the $n$ qubits of the computer) will affect the 
accuracy of the oracle implementation.  Note that the oracle acts on all 
$n$ qubits, (see Eq.~(\ref{Equation::NoisyOracle})).  It seems 
very unlikely that $\error_{rms}$ would 
decrease as $\numdim$ increases, since a greater number of qubits 
generally introduces more potential for noise.  In the best case scenario, 
it might be possible to find a physical implementation for which 
$\error_{rms}$ is essentially constant over a large range of values for 
$\numdim$.  To ensure that our analysis is as general as 
possible, we will assume here that $\error_{rms}$ scales as 
$\numdim^{-\delta}$,
where $\delta$ is some constant 
that may take both positive and negative values, or zero. We can then determine what effect the scaling 
constant $\delta$ has on algorithm performance.  In particular, we shall determine the effect of $\delta$ upon the complexity of the algorithm, {\it i.e.}, upon the number of Grover iterations required to achieve a given probability of successful measurement of the target state $\ket{1}$.  This will allow us to further determine the maximum allowable oracle error scaling that ensures a constant probability of success independent of the library size, for a given rate of library growth.

\subsection{Dynamics and scaling of the noisy Grover search algorithm}
\label{subsection:dynamics}

To analyze the dynamics of the Grover iteration, we observe that any 
pure state on the Bloch sphere can be represented in spherical 
coordinates by two 
angles, $\theta$ and $\phi$, where we define $\theta$ to be the angle 
from the $z$-axis and $\phi$ to be the 
azimuthal angle.  Here we have chosen our coordinates such that the target state $\ket{1}$ is the 
south pole ($\theta = \pi$) and the state $\ket{2}$ is the north pole ($\theta = 0$) of our Bloch 
sphere.  Because the operator $\grover$ depends upon the random 
variable $\error$, each application of $\grover$ induces a
new probability distribution over the Bloch sphere which we will 
denote by $f(\theta, \phi)$.  In 
this notation, our initial state at time $t=0$ is given by 
a delta function (i.e. a pure state) centered on $\phi = 0$, 
with $\theta = \sin^{-1}(1/\sqrt{N}) \approx 1/\sqrt{N}$.

Using this notation, we can write the probability of obtaining the 
state $\ket{1}$ upon measurement after $t$ iterations as
\begin{equation} \label{Equation::PSuccess}
\begin{array}{rcl}
P(t) &=& 
1-\frac{1}{2}\int{\int{\cos{\theta}f_t(\theta,\phi)\sin{\theta}d\theta 
d\phi}}\\
&=& \frac{1}{2}\left(1 - \ensavg{\cos{\theta}}\right)\\
&=& \frac{1}{2}\left(1 - \ensavg{z}\right).
\end{array}
\end{equation}
Let us define the complexity of our search algorithm to be the number 
of iterations required to obtain a success probability of $2/3$.  
We immediately see that if the probability of obtaining 
state $\ket{1}$ upon measurement after $t$ iterations is 
$P(t)$, then by repeating this procedure 
approximately $\frac{2}{3P(t)}$ times, we can boost the overall 
success 
probability to 
$2/3$.  It should be noted that the choice of the constant $2/3$ is 
arbitrary; in general, the choice of constant will not affect the 
complexity of the algorithm.  Thus, the overall complexity of our 
algorithm 
is $\bigo(t/P(t))$.  Furthermore, \refeqn{PSuccess} states that in 
order to succeed with some desired probability $P^*$, our probability 
distribution function 
$f(\theta, \phi)$ must be non-negligible when $\theta \geq 
\theta^*$, where
\begin{equation}
\theta^* = \cos^{-1}\left(1 - 2 P^*\right).
\end{equation}
In other words, to obtain some desired probability of success, $P^*$, 
there must be a high probability of reaching points on the 
Bloch sphere with polar angle greater than $\theta^*$.  

Having established this terminology, we will now give a phenomenological 
description of the evolution of the probability distribution over the 
Bloch sphere.  Specifically, we would like to know what dependence 
the magnitude and scaling of the error $\error_{rms}$ has on the 
maximum attainable polar angle.  

We first rewrite the effects of the Grover iterate as a function of 
polar coordinates.  In the polar coordinates defined above, the 
$z$-rotation can be written as
\begin{eqnarray} 
\label{Equation::ZRotationPhi}
\phi & \oparrow{\rot_{\hat{z}}} & \phi + \error, \\
\label{Equation::ZRotationTheta}
\theta & \oparrow{\rot_{\hat{z}}} & \theta.
\end{eqnarray}
If we momentarily neglect the $y$-axis rotations, our $z$-axis 
rotation dynamics correspond to a random walk on the variable $\phi$ 
with periodic boundary conditions.  To analyze the 
effects of the $y$-rotation, 
we then take advantage of the fact that $1/\sqrt{N}$ 
is a small quantity and expand in powers of $1/\sqrt{N}$ to obtain
\begin{eqnarray} 
\label{Equation::YRotationPhi}
\phi & \oparrow{\rot_{\hat{y}}} & \phi - 
\sin{\phi}\frac{\cos{\theta}}{\sin{\theta}}\frac{4}{\sqrt{N}} + 
\bigo(1/N), \\ 
\label{Equation::YRotationTheta}
\theta & \oparrow{\rot_{\hat{y}}} & \theta + 
\cos{\phi}\frac{4}{\sqrt{N}} + \bigo(1/N).
\end{eqnarray}
Finally, we can write the combined effects of our noisy Grover operator 
$\grover$ using Eqs.\ 
(\ref{Equation::ZRotationPhi}--\ref{Equation::YRotationTheta}), as 
\begin{eqnarray} 
\label{Equation::GRotationPhi}
\phi & \oparrow{\grover} & \phi - \sin{\phi}\frac{\cos{\theta}}{\sin{\theta}}\frac{4}{\sqrt{N}} + \error, \\ 
\label{Equation::GRotationTheta}
\theta & \oparrow{\grover} & \theta + \cos{\phi}\frac{4}{\sqrt{N}},
\end{eqnarray}
where we have dropped terms of $\bigo(1/N)$.

Having written the dynamics of the Grover operator in 
terms of polar coordinates on the Bloch sphere, we now consider the 
probability distribution of the quantum state over the Bloch sphere after 
$T$ Grover iterations.  We will analyze the dynamics for $\phi \ll 1$, and 
then consider when this approximation is valid.  In this regime, we can 
approximate the Grover operator dynamics as
\begin{eqnarray} \label{Equation::GroverSmallPhi}
\phi & \oparrow{\grover} & \phi + \error,\\
\theta & \oparrow{\grover} & \theta + \frac{4}{\sqrt{N}}.
\end{eqnarray}
It is evident that in this small $\phi$ limit, the $\phi$ dynamics are 
completely determined by $z$-axis rotation, 
which results in a random walk on the variable $\phi$ that is controlled 
by the random variable $\error$.  
The central limit theorem tells us that regardless of the probability 
distribution $\prob(\error)$ from which the random variable $\error$ 
is sampled, after an adequate number of iterations the distribution 
of $\phi$ will converge to a Gaussian with width 
$\error_{rms}\sqrt{T}$:
\begin{equation} \label{Equation::WidthPhi}
\phi_{rms} \propto \error_{rms}\sqrt{T},
\end{equation}
where $T$ is the number of iterations.

Let us now consider when this approximation is valid.  Clearly, the 
condition that $\phi \ll 1$ is satisfied for the initial state $\phi_0 = 
0$.  We recall that $\error_{rms}$ is assumed to scale as 
$N^{-\delta}$, where $\delta$ is some constant.  The validity of our small 
$\phi$ approximation is found to be highly dependent on the scaling 
exponent $\delta$.  We consider two cases: $\delta > 1/4$ and $\delta 
\leq 1/4$.  

If $\delta > 1/4$, we examine the probability distribution after $T 
= \lambda \sqrt{N}$ iterations, where $\lambda$ is some small constant.  
Inspection of \refeqn{WidthPhi} shows that 
the probability distribution of $\phi$ after $T$ steps will then have 
width $\sqrt{\lambda} N^{1/4-\delta}$. 
Since $N$ can certainly be made arbitrarily large, this justifies our 
assumption that $\phi$ is small.  Within this regime, 
\refeqn{GroverSmallPhi} shows that the dynamics of 
the variable $\theta$ are simply those of the
deterministic $y$-rotation with constant increment $\frac{4}{\sqrt{N}}$.  After $T$ iterations of this, we obtain 
\begin{equation} \label{Equation::UpperBound}
\begin{array}{rcl}
\theta_T &=& \theta_0 + \frac{4T}{\sqrt{N}}, \\
& \approx& \frac{4T}{\sqrt{N}},
\end{array}
\end{equation}
which is the same as the large $N$ limit of the noiseless Grover search, Eq.~(\ref{Equation::NoisyOracle}). 
Thus when $\delta > 1/4$, the error has no effect on algorithmic 
complexity for significantly large $N$, yielding a complexity 
of $\bigo(N^{1/2})$.  This bound is trivially tight 
since the search problem is well-known to have a lower bound of 
$\Omega(N^{1/2})$\cite{Bennett:97}.
For the remainder of 
this section, 
we will therefore analyze the complementary case, {\it i.e.}, when 
$\error_{rms}$ 
scales as $N^{-\delta}$ with $\delta \leq 1/4$, and determine the effect of 
the error on algorithmic complexity.  We note that this latter case 
includes the best-case physical situation of 
$\error_{rms}$ independent of $N$, {\em i.e.}, $\delta = 0$.

In order to analyze the system dynamics for $\delta \leq 1/4$,  
let us examine the probability distribution after $T$ iterations such that
\begin{equation} \label{Equation::BigT}
T = \lambda / {\error}_{rms}^2,
\end{equation}
where $\lambda$ is a small constant.  By \refeqn{WidthPhi}, the 
distribution width $\phi_{rms}$ is proportional to $\sqrt{\lambda}$.  
Thus, by selecting a small enough 
constant $\lambda$, our approximation that $\phi$ is small is again justified.  
Using \refeqn{GroverSmallPhi} and \refeqn{UpperBound}, we find that the $\theta$-rotation is again essentially 
deterministic and that
\begin{eqnarray} \label{Equation::UpperBound2}
\theta_T &=& \frac{4T}{\sqrt{N}}, \nonumber \\
&=& \frac{4\lambda}{\error_{rms}^2 \sqrt{N}} \nonumber \\
&=& O\left(\frac{1}{\error_{rms}^2\sqrt{N}}\right).
\end{eqnarray}
Using \refeqn{PSuccess}, we obtain
\begin{equation}
\begin{array}{rcl}
P(T) &=& 
\frac{1}{2}\left(1-\cos{\left(\frac{\lambda}{\error_{rms}^2\sqrt{N}}\right)}\right) \\
&\approx& \frac{\lambda^2}{4\error_{rms}^4 N},
\end{array}
\end{equation}
where we have used the fact that $\frac{\lambda}{\error_{rms}^2\sqrt{N}} 
\ll 1$ when $\delta \leq 1/4$.  By proving that we can attain 
(essentially 
deterministically, as described above) a polar 
rotation of \emph{at least} $\frac{4\lambda}{\error_{rms}^2\sqrt{N}}$ in 
$T = \lambda / \error_{rms}^2$ iterations, we have shown that 
the complexity of the noisy search algorithm is $\bigo(\frac{T}{P(T)}) = 
\bigo(\error_{rms}^2 N)$.  Taking the scaling of $\error_{rms}$ into 
account ($\error_{rms} \sim N^{-\delta}$), leads to the overall 
algorithmic complexity $\bigo(N^{1-2\delta})$ \footnote{It 
should be noted that this analysis applies only in the limit of large 
$N$, which is what we mean when we talk of algorithmic complexity in 
the first place.}.  So for $\delta = 1/4$ we obtain an optimal solution 
having the same speed-up as the noiseless quantum search.  For $0 < \delta < 1/4 $, we find a scaling intermediate between the noiseless quantum and classical search algorithms.  The quantum speed-up factor decreases as $\delta$ approaches zero and is completely lost when $\delta = 0$.  (Note that the quantum search formally becomes slower than the classical search in the worst case scenario when the phase errors increase with $N$, {\it i.e.}, $\delta < 0$.)

In order to show that this complexity bound is tight, we now will 
show that we can attain \emph{at most} a polar rotation 
of $\bigo\left(\frac{1}{\error_{rms}^2\sqrt{N}}\right)$
in sub-classical time.  Repeating the argument above following 
\refeqn{UpperBound2} will then lead to the identification of 
$\Omega(N^{1-2\delta})$ as a lower bound.  Thus, we find that 
$\bigo(N^{1-2\delta})$ is a 
tight 
bound on the complexity in presence of noise.  Obtaining a lower bound 
on the algorithmic 
complexity is important because without it, it 
would be unclear whether we could further reduce our search complexity by 
running the noisy search algorithm for more than $\bigo(1/ 
\error_{rms}^2)$ 
iterations.  The lower bound will demonstrate that after 
$\bigo(1/\error_{rms}^2)$ 
iterations, it is impossible to achieve a super-classical enhancement in 
success probability.

To achieve a polar rotation greater than 
$\bigo\left(\frac{1}{\error_{rms}^2\sqrt{N}}\right)$, we will need to obtain a 
significant probability density on the region of the Bloch sphere described by
$\theta \geq \frac{c}{\error_{rms}^2\sqrt{N}}$, where $c$ is a 
constant.  
Let us therefore consider the dynamics of a state on this 
region of the Bloch sphere.  
Given a pure state with $\phi_0 = 0$ and $\frac{\pi}{2} \geq \theta_0 \geq 
\frac{c}{\error_{rms}^2\sqrt{N}}$, we examine the action of $T$ 
iterations of 
the Grover algorithm on this state, where $T = \lambda / 
\error_{rms}^2$.  Now, instead of choosing $\lambda$ small as previously, 
we select $\lambda \gg 1$, which ensures a large width for the long time 
distribution of the azimuthal angle $\phi$, \refeqn{WidthPhi}.  Thus 
after $T$ iterations with this large width, we obtain a completely
uniform distribution on the variable $\phi$.  We now consider what happens
to the polar variable, $\theta$.  Referring to \refeqn{ZRotationTheta} and 
\refeqn{YRotationTheta}, we find that after $T$ iterations we obtain
\begin{equation} \label{Equation::ThetaWindow}
\begin{array}{rcl}

\abs{\theta_T - \theta_0} &\leq& \frac{T}{\sqrt{N}} \\
&\leq& \frac{\lambda}{\error_{rms}^2\sqrt{N}} \\
&=& \bigo\left(N^{-1/2+2\delta}\right).
\end{array}
\end{equation}
Again, provided that $\delta \leq 1/4$, this difference can be 
made arbitrarily small by selecting a large enough $N$.  Thus, after 
$T$ iterations, our probability distribution of the polar coordinate
$\theta$ is confined within an arbitrarily small window around $\theta_0$, 
but is uniformly distributed on the azimuthal coordinate $\phi$.  Figure 1 
shows a graphical representation of these dynamics.

Let us now consider the effects when the Grover operator is subsequently 
applied to such a probability distribution that is uniform in $\phi$.  
We will show that there exists a 
symmetry in the associated $\theta$ transformations in this regime, such 
that one half of all azimuthal angles, $\phi$, are associated with a 
$\theta$ rotation in one direction, and the other half are associated 
with $\theta$ rotations of equal magnitude in the opposite direction.  
To reveal this symmetry, we show that the 
transformation $\phi \to \phi + \pi$ on 
\refeqs{GRotationPhi}{GRotationTheta} results in an exactly opposite 
rotation in the $\theta$ direction to that associated with the azimuthal 
angle, $\phi$.  From \refeqn{ThetaWindow}, we know that the 
distribution of $\theta_T$ will be arbitrarily close to $\theta_0$.  Since 
we have assumed that our initial state was located in the region $\pi/2 
\geq \theta_0 \geq \frac{c}{\error_{rms}^2 \sqrt{N}}$, we can 
rewrite \refeqn{GRotationPhi} as follows:
\begin{eqnarray} \label{Equation::GRotation2}
\phi & \oparrow{\grover} & \phi - 
\sin{\phi}\frac{\cos{\theta}}{\sin{\theta}}\frac{1}{\sqrt{N}} + 
\error \nonumber \\ 
 & \approx & \phi - 
\sin{\phi}\frac{\error_{rms}^2 \sqrt{N}}{c}\frac{1}{\sqrt{N}} + 
\error  \nonumber \\ 
 & = & \phi - 
\sin{\phi}\frac{\error_{rms}^2}{c} + 
\error.  
\end{eqnarray}
We now make use of the arbitrariness in choice of the constant $c$, 
choosing $c \gg 1$ so that $c \gg \sin{\phi}$, making the second term 
arbitrarily small.  Hence we obtain
\begin{eqnarray} \label{Equation::GRotation3}
\phi & \oparrow{\grover} & \phi + \error,
\end{eqnarray}
which demonstrates that the transformation $\phi \to \phi+\pi$ has no net 
effect on the azimuthal dynamics.
On the other hand, our $\theta$ dynamics become
\begin{eqnarray}
\theta & \oparrow{\grover} & 
\theta + 
\cos({\phi+\pi})\frac{4}{\sqrt{N}} + \bigo(1/N) \nonumber \\
\label{Equation::GroverTransformedTheta}
&=& \theta - \cos{\phi}\frac{4}{\sqrt{N}} + \bigo(1/N).
\end{eqnarray}
\refeqn{GroverTransformedTheta} 
shows that the transformation $\phi \to 
\phi + \pi$ causes the polar angle $\theta$ to be rotated in the opposite 
direction, as a consequence of the change in 
sign of the $\cos{\phi}$ factor.  It follows that once $\phi$ has
reached a uniform distribution, there will be an equal probability of 
rotating $\theta$ by $4/\sqrt{N}$ in the positive and negative 
directions.  This behavior is valid on the entire region of the 
Bloch sphere specified by our initial condition $\pi/2 \geq 
\theta_0 \geq \frac{c}{\error_{rms}^2\sqrt{N}}$.  As 
a result, the $\theta$ dynamics over this entire region no longer resemble a 
deterministic rotation towards $\theta = \pi$, but instead resemble a random walk on 
$\theta$ with step size of approximately $4/\sqrt{N}$.  Figure 1 shows a 
graphical representation of these dynamics.

Using the central limit theorem, we find that
\begin{equation}
\theta_{rms} \propto \sqrt{t/N},
\end{equation}
where $t$ is the number of 
iterations after the uniform distribution in $\phi$ has been attained, {\it i.e.}, after the first $T$ iterations.  Consequently, to ``move'' the polar angle of the 
probability distribution by some small angle $\Delta \theta$ now takes 
time proportional to $\Delta \theta ^2 N$.  From 
\refeqn{PSuccess}, the corresponding change in the probability of 
a successful measurement of the target state is 
\begin{equation}
\Delta P \approx \frac{1}{2}\sin{\Delta\theta} \Delta\theta
\approx \frac{1}{2}\Delta\theta^2
\propto \Delta t / N
\end{equation}
Thus, once we have entered the region of the Bloch sphere characterized by 
$\pi/2 \geq \theta \geq \frac{c}{\error_{rms}^2\sqrt{N}}$, the probability 
of a successful measurement further increases only at the rate $t / N$.  This 
is the same as the classical result and consequently, we can attain 
no further speed-up over the classical algorithm once we have entered 
this regime.  The polar angle $\bigo({\frac{1}{\error_{rms}^2 \sqrt{N}}})$ 
thus constitutes an upper bound on the rotation that can be achieved in 
sub-classical time.  Hence, by the arguments given above, our bound 
$\bigo(N^{1-2\delta})$ on the complexity of the noisy quantum search 
algorithm is a tight bound. 

\subsection{Summary of discrete time noise Grover search} \label{subsection:summary_noise}

Using the above results, we can now give an accurate phenomenological description of the 
dynamics of the noisy search algorithm.  We identify the critical timescale of the 
algorithm as $t_{mixing} = 1/\error_{rms}^2$, which is the mixing 
time of the $z$-rotation induced by the random error.  For times $t 
\ll t_{mixing}$, the algorithm proceeds more or less 
deterministically, and the initial state is rotated towards the 
target state $\ket{1}$, attaining a polar angle of $\theta = 
\frac{1}{\error_{rms}^2\sqrt{N}}$.  However, for times $t \gg 
t_{mixing}$, 
the $\phi$ variable becomes completely mixed and the $\theta$-rotation, which is the 
driving force of the search process, has 
an equal probability of increasing or decreasing the polar angle of 
the probability distribution.  As a result, the $\theta$ dynamics 
also become those of a random walk, and the searching rate becomes 
classical.  Therefore, in order to gain the maximum speed-up over the classical 
algorithm, the best methodology we can employ is to 
run 
the algorithm for $t \ll t_{mixing}$, measure, and then repeat 
this process to boost our overall success probability.  This protocol will 
allow the sub-classical scaling to be retained.

The effects of the scaling of  
$\error_{rms}$ on the complexity of the search algorithm are seen to be different according to the value of the error scaling index $\delta$.  We 
saw that if $\delta > 1/4$, the search algorithm was 
essentially unaffected by the presence of error and the complexity is 
identical to the noiseless quantum search result of 
$\bigo(\sqrt{N})$.  On the other hand, if $\delta \leq 1/4$, then the 
optimal speed-up is obtained by running the algorithm for $t = 
\bigo(1/\error_{rms}^2)$ iterations, achieving a 
maximum polar angle of $\theta = \bigo(1/\error_{rms}^2\sqrt{N})$ 
and then measuring.  Expanding
\refeqn{PSuccess} yields the measurement 
success rate, $P(t) = \bigo(1/\error_{rms}^4 N)$ and a resulting algorithmic 
complexity of $\bigo(t/P(t)) = \bigo(\error_{rms}^2 N) = 
\bigo(N^{1-2\delta})$.  
So the optimal error scaling is given by $\delta 
\geq 1/4$, {\em i.e.}, the errors scale as $\bigo(N^{-1/4})$.  In 
contrast, the most physically realistic 
constant error scaling $\delta=0$, corresponding to a constant error over 
a range of $N$ values, yields no speed up over the classical search.

It is interesting to
compare these results to the analytic results for a constant phase error
oracle given in \cite{Long:99}.  In the constant phase error model, the
oracle applies the same phase $\exp{(i(\pi+\epsilon))}$ to the marked 
state
at each iteration.  It can be shown that for a constant phase
error of magnitude $\epsilon$, the error must scale as 
$\epsilon = \bigo(\numdim^{-1/2})$ in
order to obtain a quadratic speed-up over the classical search.  The
differences in error accumulation between the constant and random phase
error processes can be compared (for example) to the ballistic and
diffusive regimes of Brownian motion, respectively.  Constant phase errors 
tend to
accumulate quickly since they are all in the same ``direction''.  On the
other hand, subsequent random phase errors can cancel each other out and
thus accumulate more slowly.

Let us now consider a question of physical importance.  Given an oracle with error 
magnitude $\error_{rms}$ and a library of size $N$, let us assume that we can 
attain some constant success rate $P$ after $T$ iterations.  Given a larger library 
of size $N' = k N$, we wish to obtain the same success rate $P$ after 
$T' = \sqrt{k}T$ iterations.  In other words, we wish to obtain the ideal 
quantum search complexity of $O\left(\sqrt{N'}\right)$ also for a larger library size.  What is the maximum 
allowable oracle error, $\error_{rms}'$, for the larger library?  The 
answer to this question follows immediately from our previous analysis.  
We know that $\error_{rms}$ can scale as $N^{-\delta}$ with $\delta \geq 1/4$ 
without affecting algorithmic complexity.  Thus the maximum allowable 
error in the oracle for the larger library is $\error_{rms}' = 
k^{-1/4}\error_{rms}$.  As an example, if the library size is doubled, the oracle error for the larger library is required to be at most 0.84 the corresponding error for the smaller library.

To verify the validity of our analytic results, we can simulate the effects of a 
noisy oracle numerically.  Given a library of size $N$, we simulated the Grover's 
search algorithm using the procedure described in Section~\ref{subsection:noisy}.  For each iteration of the search algorithm, 
a phase error $\error$ was selected from a zero-mean Gaussian distribution with 
standard deviation $\error_{rms}$.  The algorithm was run for $t = \pi \sqrt{N} / 
4$ iterations and the maximal probability of success attained was recorded.  This 
process was repeated $100$ times and the average success probability was 
calculated.  Figure 2 
plots the average probability of success, $\ensavg{P}$, versus the library size, 
$n=\log{N}$, for various values of $\error_{rms}$.  Figure 3 shows the value of 
$\log{\error_{rms}}$ that yields a constant success probability of $P=1/2$ for 
a given library size.  As predicted by our analysis above, we find that 
in order to obtain a 
constant success rate as a function of $n$ (or $N$), the error must scale as $\error_{rms} \propto N^{-1/4}$.

\section{Effect of Noise on Continuous Time Analogue of Quantum Search} \label{Section::ContinuousTime}

In this section, we analyze the effect of noise on a slightly different model for quantum search that has been proposed 
by Farhi and Gutmann \cite{Farhi:98}.  In this model, one again starts in the symmetric
superposition of all states, $\left| \eta \right\rangle $, and then applies
the following Hamiltonian
\begin{equation}
H_0=|\eta \rangle \langle \eta |+|\tau \rangle \langle \tau |=H_{\eta
}+H_{\tau },
\label{Equation::GF}
\end{equation}
where $\left| \tau \right\rangle $ is the marked state. \ Note that this
Hamiltonian is directly related to the Grover's iterate, \refeqs{iterate}{Oracle}.   Action of \refeqn{iterate} amounts to 
simply
applying the Hamiltonian $H_{\eta }$ for a time $\pi $, followed by applying
the Hamiltonian $H_{\tau }$ for a time $\pi .$ \ Clearly the two operators 
\refeqn{iterate} and  \refeqn{GF} would be
equivalent if  $H_{\eta }$ and $H_{\tau }$ commuted. $\ $However, $\left[
H_{\eta },H_{\tau }\right] =\frac{1}{\sqrt{N}}(|\eta \rangle \langle \tau
|+|\tau \rangle \langle \eta |)$.  Thus the two methods become similar for large $N$ and are formally
equivalent as $N$ approaches infinity.

Farhi and Gutmann calculated the time evolution of the system when one starts in the equal superposition state $|\eta\rangle$.  They found that
\begin{equation}
e^{-iHt}\left| \eta \right\rangle
= 
e^{-it}\left(\left(\frac{1}{\sqrt{\numdim}}\cos{\left(\frac{t}{\sqrt{N}}\right)} - 
i\sin{\left(\frac{t}{\sqrt{\numdim}}\right)}\right)\ket{\tau} + 
\sqrt{\left(1-\frac{1}{\numdim}\right)}\cos{\left(\frac{t}{\sqrt{\numdim}}\right)}\ket{\eta}\right).
\end{equation}

They noticed that at time  $t=\frac{\pi \sqrt{N}}{2}$, the initial state, $|\eta\rangle$,
has evolved to the marked state, $\ket{\tau} = \ket{1}$. \ The time required to evolve to the marked state scales as $O(\sqrt{N})$, matching the complexity of Grover's algorithm with respect to an oracle. As a result, we take the time 
that it takes to reach the marked state as a function of $N$ to be the
measure of the complexity of a continuous time algorithm.

As noted above, in the limit of large $N$ the continuous time and discrete Grover's algorithm are formally equivalent. Therefore, it seems useful to also evaluate the effect of a fluctuating
''oracle'' in the continuous time picture. \ The continuous time Hamiltonian
that models a discrete quantum search noisy oracle with phase fluctuation 
$%
\epsilon $ is given by

\begin{eqnarray} \label{Equation::H_fluctuate}
H & = & H_{\eta}+(1+\xi )H_{\tau } \\ \nonumber
 & = & |\eta \rangle \langle \eta |+(1+\xi )|\tau\rangle \langle \tau|.
\end{eqnarray}
Here $\xi $ is a time-dependent, Markovian stochastic variable that 
satisfies $\int_{0}^{\pi }\xi dt=\epsilon$.  
We shall assume that $\epsilon$ fluctuates and that it can be described 
by a Gaussian distribution.

In order to evaluate the Hamiltonian and its effect on the initial
state, it is simpler to transform to our orthonormal two state basis 
$\ket{1}$ and $\ket{2}$ defined by \refeqs{BasisKetA}{BasisKetB}.
One can then decompose the transformed Hamiltonian into the corresponding 
spin operators to find that
\begin{equation}
H=\left( 1+\frac{\xi }{2}\right) {\bf I}+\left( 
\frac{1}{N}+\frac{\xi}{2}\right) \sigma _{z}+\frac{1}{\sqrt{N}}\sqrt{\left( 1-\frac{1}{N}\right) }%
\sigma _{x}.
\end{equation}

When the fluctuations in the time evolution operator $e^{-iHt}$ are now 
considered, {\it i.e.}, the resulting phase $\epsilon \neq 0$, it is 
useful to approach the problem by
examining the evolution of the density matrix.  Consider an initial 
density matrix,
\begin{equation}
\rho(0) = \left[
\begin{array}{cc}
a & b \\
b^{\ast} & d 
\end{array}
\right].
\end{equation}
It is well established that a 
fluctuating $\sigma _{z}$ component of the Hamiltonian leads to
dephasing. \ For example, applying the fluctuating perturbation $\frac{\xi}{2}\sigma
_{z}$ for a time $\pi $ yields,
\begin{equation}
\rho(t) = \exp{\left(-i\sigma_z \pi/2\right)}\rho(0)\exp{\left(i\sigma_z 
\pi/2\right)}.
\end{equation}
Subsequently averaging over all possible values 
of the resulting total phase, $\epsilon$, leads to the following 
evolution:
\begin{equation}
\int \frac{1}{\sqrt{2\pi 
}s}e^{-{\epsilon^2}/{2s^2}}e^{-i\epsilon\sigma_{z}/2}\left[ 
\begin{array}{cc}
a & b \\ 
b^{\ast } & d
\end{array}
\right] e^{i\epsilon^2/{\sigma _{z}}}d\epsilon =\left[ 
\begin{array}{cc}
a & be^{-s^2/2} \\ 
b^{\ast }e^{-s^2/2} & d
\end{array}
\right]. 
\end{equation}
Here $s^{2}$ is the variance of the fluctuating phase $\epsilon$.
Since it has been assumed that the fluctuating field $\xi$ is 
Markovian, 
one can treat this as
a dephasing term described by a decay constant $\Gamma = s^2/{2\pi}$
and then use the corresponding Bloch equations~\cite{Blum:81} to calculate 
the evolution of the
system. \ This Markovian approximation is valid when the fluctuation of $\xi$ is much faster than $\frac{1}{\sqrt{N}}$. 
To analyze the system evolution, we therefore decompose our density matrix using the Bloch representation, 
\begin{equation} \label{Equation::bloch}
\rho =\frac{1}{2}{\bf I}+\frac{1}{2}\hat{n}\cdot {\bf \sigma }.
\end{equation}
The time evolution is calculated by
solving the Bloch equations for the components of $\hat{n}$ 
\cite{Blum:81}: \ 
\begin{eqnarray}
\stackrel{.}{n}_{x} &=&\frac{2}{N}n_{y}-\Gamma n_{x},  \\
\stackrel{.}{n}_{y} &=&\frac{2}{\sqrt{N}}\sqrt{\left( 1-\frac{1}{N}\right) 
}n_{z}-\frac{2}{N}n_{x}-\Gamma n_{y},  \\
\stackrel{.}{n}_{z} &=&-\frac{2}{\sqrt{N}}\sqrt{\left( 
1-\frac{1}{N}\right) }n_{y}.
\end{eqnarray}

In order to understand the effect of the dephasing on the algorithmic complexity, we 
examine these equations in the limit of large $N$ and keep only terms that are
of order $\frac{1}{\sqrt{N}}.$ \ This yields only two coupled equations, in $n_y$ and in $n_z$:

\begin{eqnarray}
\stackrel{.}{n}_{y} &=&\frac{2}{\sqrt{N}}n_{z}-\Gamma n_{y} \\
\stackrel{.}{n}_{z} &=&-\frac{2}{\sqrt{N}}n_{y}.
\label{Equation::Bloch_2}
\end{eqnarray}

Our initial density matrix is given by
\begin{eqnarray} \label{Equation:initial_rho}
\rho & = & |\eta \rangle \langle \eta | \nonumber \\
     & = & \frac{1}{2}
{\bf I}+\frac{1}{2}\left(-1+\frac{2}{N}\right)\sigma_{z}+\frac{1}{2}\left(\frac{2\sqrt{N-1}
}{N}\right)\sigma_{x}.
\end{eqnarray}  
We note that the quantity $n_{z}$ is the projection 
onto 
the state $\ket{1}$ and thus provides a measure of how well the computation is
proceeding. \ Initially we have $n_{z}=-1+\frac{2}{N}$, while our computation becomes
complete when $n_{z}=1.$ \ Solving the differential equations in the 
large $N$
limit, (\refeqn{Bloch_2}), with initial conditions 
$n_{y}(0)=0$, $n_{z}(0)=-1+\frac{2}{N}$, yields the following 
solution:

\begin{equation}
\begin{array}{llll}
n_{y}(t) &=&\frac{2(-1+\frac{2}{N})}{\sqrt{\Gamma^2N-16}}&
\bigg(\exp{\left(\frac{t}{2}\left(-\sqrt{\Gamma^2}-\sqrt{\Gamma^2-16/N}\right)\right)}+\exp{\left(\frac{t}{2}\left(-\sqrt{\Gamma^2}-\sqrt{\Gamma^2-16/N}\right)\right)}\bigg)  
\\
n_{z}(t) 
&=&\frac{(-1+\frac{2}{N})}{2\sqrt{\Gamma^2N-16}}&
\bigg(\left(-\sqrt{\Gamma^2N}+\sqrt{\Gamma^2N-16}\right)\exp{\left(\frac{t}{2}\left(-\sqrt{\Gamma^2}-\sqrt{\Gamma^2-16/N}\right)\right)} 
\\ 
& & & + \left(\sqrt{\Gamma^2N}+
\sqrt{\Gamma^2N-16}\right)\exp{\left(\frac{t}{2}\left(-\sqrt{\Gamma^2}+\sqrt{\Gamma^2-16/N}\right)\right)}\bigg). 
\label{Equation::dynamics}
\end{array}
\end{equation}

Ideally, we would like to find the time at which $n_z(t)=1$, or equivalently,
the time when the probability of reaching the marked state is unity,
\begin{equation} \label{Equation::Pt}
P(t)=\langle \tau|\rho(t)|\tau \rangle=\frac{1}{2}\big(1+n_z(t)\big)=1.
\end{equation}
When measuring the complexity of Grover's algorithm however, we need only to find 
the time required such that the probability $P(t)$ of being in the marked 
state $\tau$ is greater than some constant.  For concreteness, we choose 
here the minimum time satisfying $P(t)=\langle\tau |\rho(t) |\tau \rangle 
\geq 1/4 $.

To determine this time explicitly, we calculate $P(t)$ as a function of various 
values of $\Gamma$ in the limit of large $N$.  We observe that in 
\refeqn{dynamics} the term $\sqrt{\Gamma^2-16/N}$ is imaginary when 
$\Gamma<\frac{4}{\sqrt{N}}$.  Consequently, we expect that the behavior 
will be drastically different for the two regimes a) $\Gamma < 
\frac{4}{\sqrt{N}}$, and b) $\Gamma > \frac{4}{\sqrt{N}}$. In order to 
make a direct comparison between 
the continuous time behavior and the discrete time results, we choose the 
dephasing constant $\Gamma$ to scale with $N$ in the same way as in Section \ref{Section::DiscreteTime},  namely
\begin{equation} \label{Equation::GammaScaling}
\Gamma=\alpha N^{-2\delta}\propto \epsilon^2_{rms},
\end{equation}
where $\alpha$ is a proportionality constant. 
This provides the contact point of the continuous time search algorithm with
the discrete time algorithm of Section~\ref{Section::DiscreteTime}.
We then see that regimes a) and b) correspond to the two regimes already 
established in Section~\ref{Section::DiscreteTime}, {\it i.e.}, in region 
a), $\delta \geq 1/4$, and in region b), $\delta \leq 1/4$.  We note that 
our regions b) and a) both include $\delta=1/4$.  The boundary line 
between the regions occurs when $\delta=1/4$ and $\alpha=4$.

For $\Gamma=\alpha N^{-2\delta}<\frac{4}{\sqrt{N}}$, {\it i.e.}, in region 
a), we calculate from \refeqn{dynamics} that
\begin{equation}
P(t)=\frac{1}{2}+\frac{1}{2}\left(-1+\frac{2}{N}\right)e^{t/{2\Gamma}}\left(\cos(\sqrt{16/N-\Gamma^2}t/2)+i\sqrt{\frac{\Gamma^2N}{\Gamma^2N-16}}\sin(\sqrt{16/N-\Gamma^2}t/2)\right).
\end{equation}
We now pick a time 
\begin{equation} \label{Equation::bigd}
t'=\frac{2\pi}{\sqrt{16/N-\Gamma^2}}
\end{equation}
and find that
\begin{eqnarray}\label{Equation::subs}
P(t')&=& 
\frac{1}{2}+\frac{1}{2}\left(1-\frac{2}{N}\right)\exp\left(-\frac{\pi\Gamma}{\sqrt{16/N-\Gamma^2}}\right).
\end{eqnarray}
Hence,
\begin{eqnarray}
P(t')&>&\frac{1}{2}-O(1/N).
\end{eqnarray}
Since this is larger than the value corresponding to our definition of 
minimum time, it implies that the complexity of the search algorithm is 
bounded from above by $O(t')$.  Inspection of \refeqn{bigd} shows that 
$t'$ is an increasing function of $\Gamma$. Therefore, in order to 
evaluate an upper bound for $t'$ in the regime a) where $\delta \geq 1/4$, 
we need to evaluate $t'$ for the largest possible value of $\Gamma$. The 
largest value of $\Gamma$ in this regime lies on the boundary with regime 
b), namely where $\delta=1/4$ and $\alpha = 4$. Hence, we evaluate $t'$ as 
one asymptotically approaches the boundary between regions a) and b), for 
a given error scaling $\delta$. We choose $\delta=1/4$ and $\alpha=4-m$, 
where m is a small constant greater than zero. Substituting in 
\refeqn{subs} yields $t'=\frac{2\pi \sqrt{N}}{m}$. Thus the minimum time 
will be on the order of $\bigo(\sqrt{N})$, corresponding to an
upper bound on the 
algorithmic complexity of $\bigo(\sqrt{N})$.  Hence for $\delta \geq 1/4$, 
in regime a), the continuous time search algorithm achieves its maximal 
algorithmic quantum speed up.  This is in agreement with the results from 
Section \ref{Section::DiscreteTime}, which showed that the discrete time quantum search algorithm yields the maximum algorithmic speedup when the root mean square error $\error_{rms}$ is smaller than $N^{-1/4}$, corresponding to $\delta \geq 1/4$. 

To complete the continuous time analysis, we solve for $P(t)$ in the 
regime b) where $\delta\leq 1/4$. Here evaluation of $n_z$ in the limit 
of large $N$ leads to 
\begin{eqnarray}
P(t)=\frac{1}{2}-\frac{1}{2}e^{-4t/{N\Gamma}}.
\end{eqnarray}

Setting $P(t')=1/4$ and solving for $t'$ yields
\begin{eqnarray} 
t' &=& \frac{N\Gamma \ln (2)}{4}\\
&=&\bigo (N^{1-2\delta}).\label{Equation::littled}
\end{eqnarray}

We have verified our conclusions by numerically simulating \refeqn{dynamics} for various values of $N$ and $\Gamma$.  Figure 4 shows a 
logarithmic plot of the minimum time $t'$ required to obtain a success 
probability of $P(t') = 1/4$, as a function of the error scaling 
parameter $\delta$, for various 
values of $N$.  It is evident that the algorithmic complexity shows a 
marked transition at $\delta =1/4$, from scaling $\bigo(N^{1/2})$ for 
$\delta 
\geq 1/4$, to $\bigo(N^{1-2\delta})$ for $\delta \leq 1/4$, as predicted 
by \refeqn{bigd} and \refeqn{littled} respectively.  
Equivalently,  we can state that the continuous time quantum search 
algorithm with 
randomized phase error achieves minimal complexity and hence maximum 
algorithmic speedup when $\epsilon_{rms}\leq 
N^{1/4}$ ($\delta \geq 1/4$). This agrees with the results for the 
discrete-time algorithm.


\section{Conclusions} \label{Section::Conclusions}
The analysis in this paper has provided a phenomonological description of 
the dependence of the algorithmic complexity of Grover's algorithm on the 
scaling of oracle phase error for a discrete quantum search, and on stochastic Hamiltonian errors leading to phase error in a continuous time quantum search algorithm.  In both the discrete- and 
continuous-time versions of the algorithm, 
it was found that if the phase error scaled with size as $N^{-\delta}$, then for 
$\delta \leq 1/4$ the 
effect on the complexity of the algorithm for large $N$ was negligible.  
However, if the size scaling of the error lies in the regime
$\delta \leq 1/4$, then for large $N$ it was determined that there is 
tight bound of $\bigo\left(N^{1-2\delta}\right)$ on the complexity of the 
search algorithm.  In particular, this 
implies that in the presence 
of any constant (non-zero) amount of phase error in the oracle ($\delta =0$), there exists some 
library size $\maxdim$ above which the quantum search algorithm no longer 
provides a quadratic speed-up.  
In this case, for databases of size $\numdim > \maxdim$, the search time is 
$\Omega(\numdim)$, which is equivalent to the classical result and there 
is 
therefore no quantum speedup.  Intermediate error scaling, $0 < \delta < 1/4$, provides speedup intermediate between the classical and quantum limits, respectively.  These results hold for 
both the discrete-time and continuous-time quantum search algorithms, and assume
very little about the specific form of the underlying error processes.  

The complexity analysis we have made here is also important for determining the precision needed in 
scaling up a quantum search.  
For 
instance, let us assume that we are able to implement a quantum search for a library 
of size $N$ with an oracle error of magnitude $\error_{rms}$.  Then to 
perform a quantum search on a library of size $kN$ with equivalent 
{\em accuracy}, our results imply that we need to 
implement an oracle with an error of at most $\error_{rms}/k^{1/4}$.
Since this must lie in the regime $\delta \geq 1/4$, physically, this requires a system where the phase error decreases as a 
function of database size. Consequently, the precision must increase 
exponentially as a function of the number of qubits, putting severe demands on the physical realization.  In contrast, if the error and hence the precision is 
constant in the system size, ({\it e.g}., a system where the natural line 
width is independent of the number of states), then there always exists a 
database size such that the quantum approach offers no speed up over the 
classical search algorithm.   

The main consequence of the non-robustness of these forms of quantum searches to oracle 
noise that was demonstrated here is a practical limitation on the size of the library on which may 
be searched with a quadratic speed-up using quantum search algorithms without any explicit error correction.  This result has significant consequences for physical implementation of quantum search algorithms, since although quantum 
error correction can be used to reduce the error present in the oracle, such 
error correction procedures can require significant resources~\cite{Preskill:98}.  In practice it will therefore
be necessary to balance the {\em cost} of error correction (in both spatial and temporal resources) with the extent of speed-up attained by a noisy 
quantum search.  The analytic results presented in this paper provide a useful bound on
the maximum oracle error permissible if a quadratic speedup is to be retained. Above this maximum allowable error, we 
must use error correction in any physical implementation. 
Conversely, below this maximum error, we can be confident that error 
correction will not be necessary, provided that oracle phase error 
is the primary source of error.  

As a final comment we point out that although Grover's algorithm and its continuous time analogue are not inherently robust to phase error in the 
oracle, it is not clear whether other implementations of quantum search may be inherently 
robust.  Exploration of both active and passive error correction schemes for 
Grover's algorithm will therefore be a valuable direction for future work.
        
\vspace{1in}
\pagebreak[4]
\begin{figure}
\includegraphics{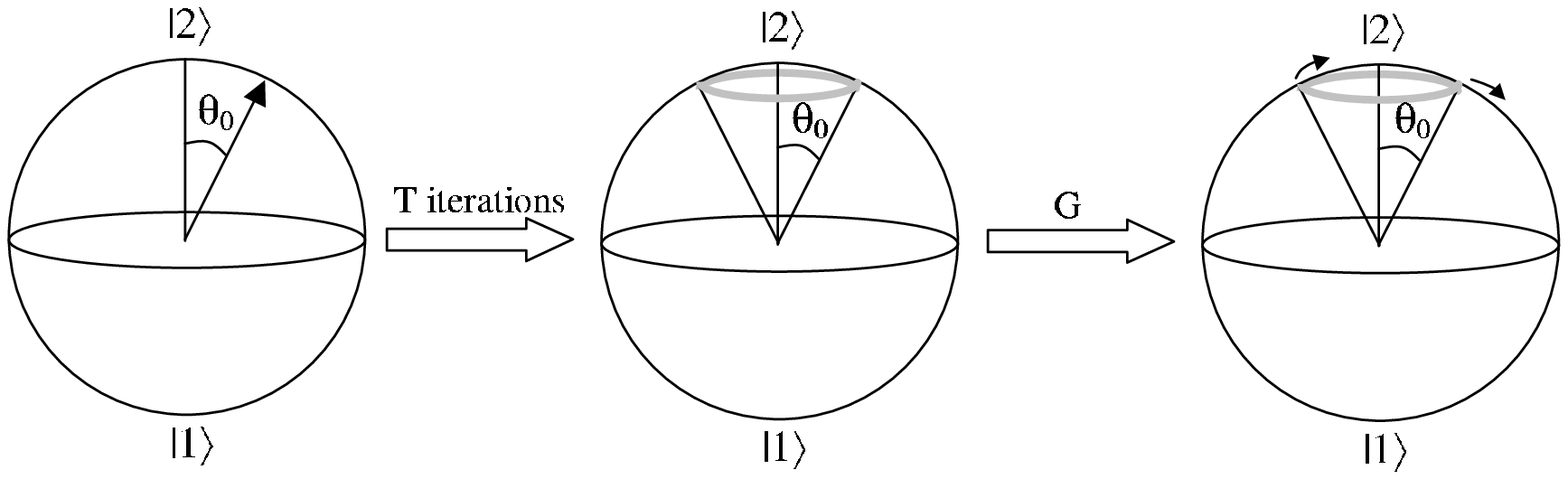}
\caption{  
Graphical representation of the noisy oracle Grover algorithm dynamics 
when $\frac{\pi}{2} 
\geq 
\theta \geq \frac{c}{{\error_{rms}}^2\sqrt{N}}$.
The target state $\ket{1}$ is located at the south pole of the Bloch sphere.  The north pole is state $\ket{2}$, \refeqn{BasisKetB}.
After $T$ iterations, a pure state at $\theta_0$ and $\phi_0$ will become 
completely mixed with respect to the $\phi$ variable but will be confined 
to an arbitrarily small window around $\theta_0$ in the $\theta$ variable.  
Subsequent applications of $G$ will then give an equal probability of 
increasing or decreasing the polar angle, corresponding to a random walk 
in $\theta$.
}

\end{figure}  

\pagebreak[4]
\begin{figure}
\includegraphics{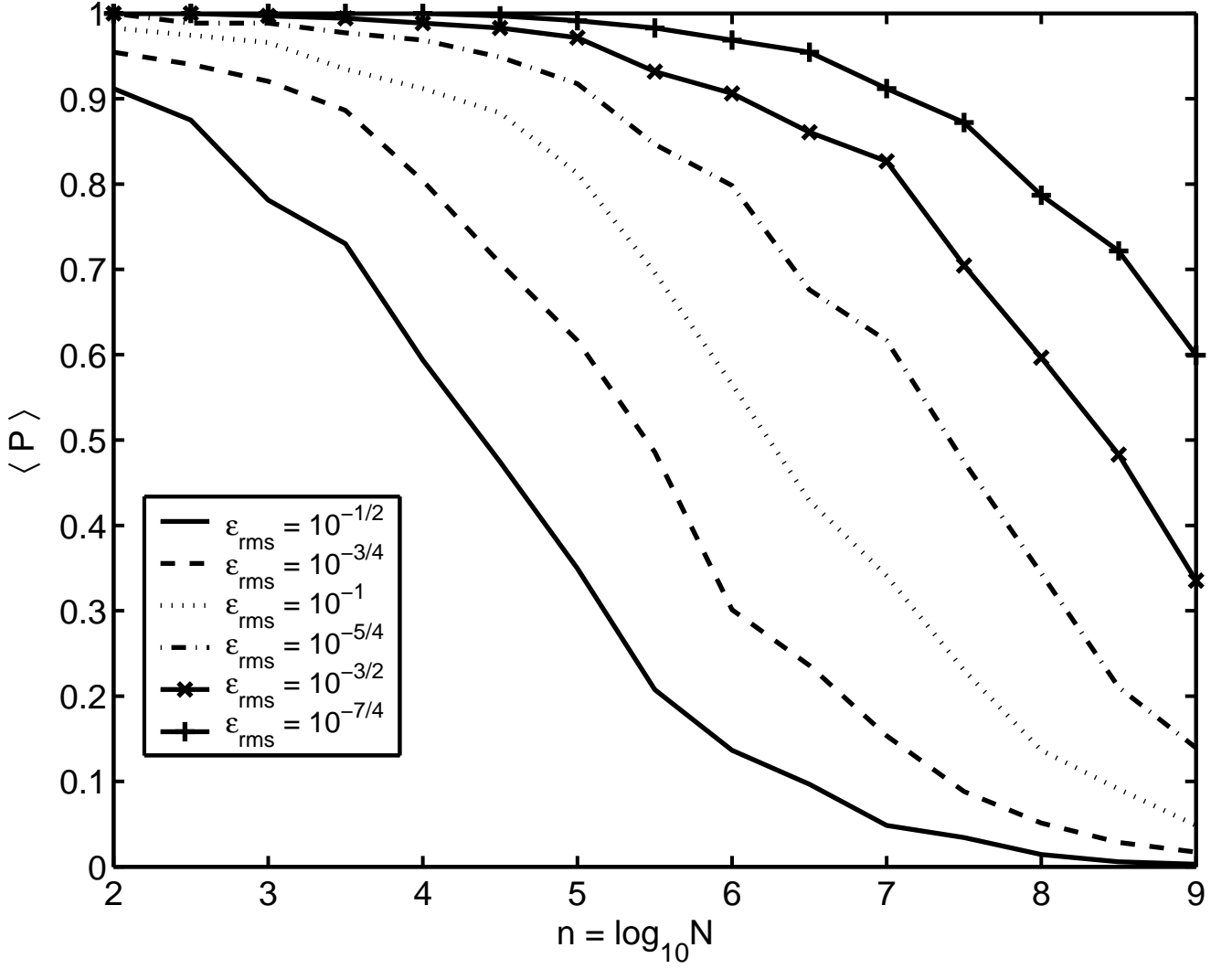}
\caption{Results of numerical simulations of the discrete time implementation of Grover's algorithm with a noisy oracle.  The oracle noise is determined by the probability distribution $p(\epsilon)$ (Section~\ref{subsection:noisy}). $\ensavg{P}$ is the  
average success rate of $100$ trials and $N$ is the library size.  The curves correspond to calculations with error magnitudes of $\error_{rms} = 
10^{-.5},10^{-.75},10^{-1},10^{-1.25},10^{-1.5},10^{-1.75}$, where the labeling goes from left to right.
}
\end{figure}  

\pagebreak[4]
\begin{figure}
\includegraphics{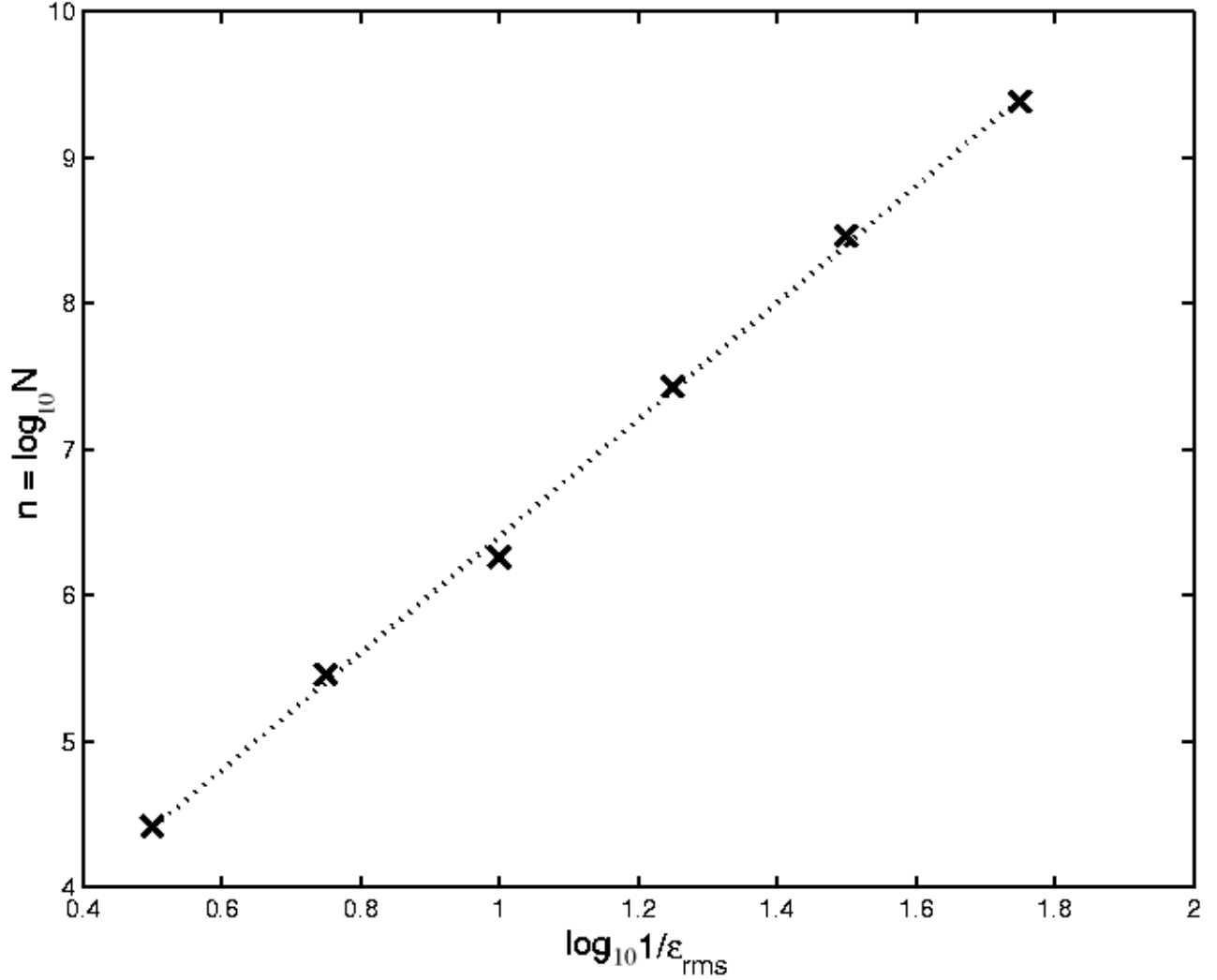}
\caption{  
Oracle error magnitude $\error_{rms}$ for given library size $N$ which yields an 
average success rate of $P=1/2$ for the discrete time implementation of Grover's algorithm with a a noisy oracle.  The slope of the best-fit line is $4.002$, which 
corresponds to $\error_{rms} \propto N^{-1/4}$ and an error scaling parameter 
$\delta = 1/4$. 
}
\end{figure}  

\pagebreak[4]
\begin{figure}
\includegraphics{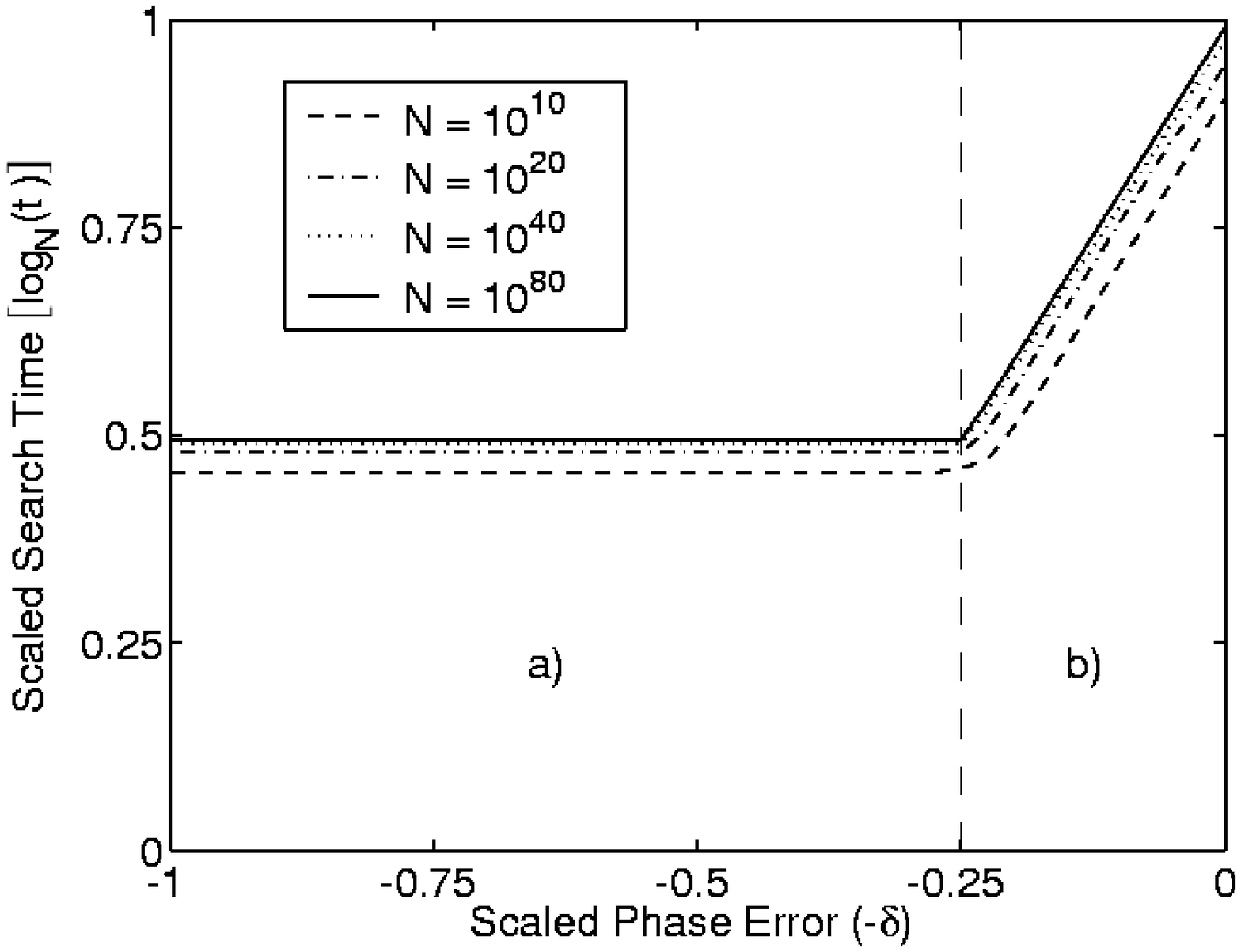}
\caption{  
Search time $t'$ for the continuous time search algorithm, shown as a 
function of the oracle phase error scaling parameter $-\delta$.  Here $t'$ 
is defined as the time it takes to achieve a success probability of $P(t') 
= 1/4$.  The oracle phase error is measured here by its size scaling 
 parameter $\delta$, where $\epsilon_{rms} = N^{-\delta}$ (see text).  The 
search time shows a distinct 
transition between two regimes a) and b).  In regime a) the continuous 
time algorithm for a database of size $N$ matches the Grover bound of $\bigo (N^{1/2})$ for large N, {\it i.e.}, $\log_N(t')=0.5$.  In regime b) the quantum search speed up is gradually lost as $\delta$ decreases from the critical value 1/4 to 0.  At $\delta =0$ we have constant error, independent of the database size, and the search time has now increased to equal the classical bound $\bigo (N)$, {\it i.e.}, $\log_N(t')=1$. In the limit of large N, the scaling of the search time with the error parameter for $\delta \geq 1/4$ (region a)) is a constant, $t'=O(N^{1/2})$, while for $\delta \leq 1/4$ (region b)), it is $t' = O(N^{1-2\delta})$ (see text).
}
\end{figure}  

\begin{acknowledgments}
NS thanks the University of California, Berkeley, for a 
Berkeleyan Fellowship.  The work of KRB was supported by the Fannie and John Hertz Foundation.  KBW thanks the Miller Foundation for Basic Research for a Miller Research Professorship 2002-2003. This effort is sponsored by the Defense Advanced
Research Projects Agency (DARPA) and the Air Force Laboratory, Air
Force Material Command, USAF, under agreement number F30602-01-2-
0524. We also thank NSF ITR/SY award 0121555.
\end{acknowledgments}

\bibliographystyle{apsrev}

\pagebreak[4]

\end{document}